\documentclass[fleqn,10pt]{wlscirep}
\usepackage[utf8]{inputenc}
\usepackage[T1]{fontenc}
\usepackage{cite}
\usepackage{amsmath,amssymb,amsfonts}
\usepackage{graphicx}
\usepackage{subcaption}
\usepackage{textcomp}
\usepackage{xcolor}
\usepackage{booktabs}
\usepackage{multirow}
\usepackage{tabularx}
\usepackage{fancyhdr}
\usepackage{tikz}

\usepackage[ruled,vlined]{algorithm2e}

\title{E-DPNCT: An Enhanced Attack Resilient Differential Privacy Model For Smart Grids Using Split Noise Cancellation\\}

\author[1,*]{Khadija Hafeez}
\author[1]{Donna O'Shea}
\author[2]{Thomas Newe}
\author[1]{Mubashir Husain Rehmani}
\affil[1]{Munster Technological University (MTU), Cork, Ireland }
\affil[2]{University of Limerick (UL), Ireland}

\affil[*]{khadija.hafeez@mycit.ie}

\begin{abstract}
High frequency reporting of energy consumption data in smart grids can be used to infer sensitive information regarding the consumer's life style and poses serious security and privacy threats. Differential privacy (DP) based privacy models for smart grids ensure privacy when analysing energy consumption data for billing and load monitoring. However, DP models for smart grids are vulnerable to collusion attack where an adversary colludes with malicious smart meters and un-trusted aggregator in order to get private information from other smart meters. {We first show the vulnerability of DP based privacy model for smart grids against collusion attacks to establish the need of a collusion resistant model privacy model. Then, we propose an Enhanced Differential Private Noise Cancellation Model for Load Monitoring and Billing for Smart Meters (E-DPNCT) which not only provides resistance against collusion attacks but also protects the privacy of the smart grid data while providing accurate billing and load monitoring. We use differential privacy with a split noise cancellation protocol with multiple master smart meters (MSMs) to achieve colluison resistance. }We propose an Enhanced Differential Private Noise Cancellation Model for Load Monitoring and Billing for Smart Meters (E-DPNCT) to protect the privacy of the smart grid data using a split noise cancellation protocol with multiple master smart meters (MSMs) to provide accurate billing and load monitoring and resistance against collusion attacks. We did extensive comparison of our E-DPNCT model with state of the art attack resistant privacy preserving models such as EPIC for collusion attack. We simulate our E-DPNCT model with real time data which shows significant improvement in privacy attack scenarios. Further, we analyze the impact of selecting different sensitivity parameters for calibrating DP noise over the privacy of customer electricity profile and accuracy of electricity data aggregation such as load monitoring and billing.
\end{abstract}
\begin{document}
\include{pythonlisting}
\flushbottom
\maketitle

\thispagestyle{empty}

\section{Introduction}

Over the past number of decades, the electric grid has been modernized, becoming more decarbonized, distributed and digitalized. Consequently, modern day electric grid systems have evolved to become smart grids allowing: two-way flow of electricity and data enabling applications such as smart metering. While smart meters provide benefits to consumers through better tracking and use of energy, more accurate billing and increased tariff options, they have also brought concerns related to privacy and data integrity over the use of personal data collected. Over the past number of years various privacy preserving techniques have been proposed to address this concern to prevent the invasion of privacy by smart meters, which include cryptography and data perturbation methods \cite{alotaibi2020comprehensive,inferPrivLeak}. 

To date much of the literature has focused on analysing the benefits of increased complexity and computation introduced through cryptography-based encryption methods versus the trade-off between privacy and utility introduced by data perturbation techniques such as Differential Privacy (DP). In addition, several works have accessed the robustness of these techniques against privacy attacks such as data reconstruction, linking, inference, differencing and correlation attacks \cite{wang2020internalAttacks}. While all these attacks differ, the goal of the adversary is to gain knowledge that was not intended to be shared. Such knowledge can be related to the data usage allowing an adversary to identify patterns and behaviours and to infer sensitive information form it. 

Chamikara et al \cite{CHAMIKARA2021192} outlined how data perturbation techniques are vulnerable to specific data reconstruction attacks such as naïve estimation, independent component analysis (ICA) and Input/Output (I/O) attacks such as eigen analysis, distribution analysis attacks and spectral filtering. The goal of all these attacks is that the adversary attempts to reconstruct the original data from perturbed data. Setting a strong perturbation has been proven to be effective against these types of attacks in advance adversarial environments. Other data perturbation attacks have focused on removing the level of noise on masked data such as Filtering Attack \cite{Barbosa2016} and Negative Noise Reduction \cite{Hale2019} attacks, which used in combination with other attacks could increase their efficacy. 
\subsection{Purpose}
To date there has been no evaluation on how the privacy models that uses pure perturbation techniques such as Differential Privacy (DP) are resistant to collusion attacks. In this specific type of attack, a group of smart meters and/or (third party) aggregators collectively work together to leak sensitive information with the aim of reconstructing private data or injecting false packets with the aim of modifying the integrity of the data sent to the utility provider. The use of trusted third party aggregators in smart grid systems make it particularly vulnerable to these types of data reconstruction/privacy and integrity attacks. The existing privacy solutions for smart grids that are collusion resistant either uses hybrid (DP with encryption) or pure encryption based solutions \cite{dream,baza2021privacy,epic,wang2020internalAttacks,Zhang,Mustafa}, which have high computation and communication cost. 

Given the above context, in this paper we present a collusion resistant Enhanced Differential Privacy with Noise Cancellation Technique (E-DPNCT) scheme, that not only preserves the privacy of smart meter, but also protects the data from being reconstructed by colluding entities such as smart meters and trusted third party aggregators. E-DPNCT extends previous work, DPNCT\cite{hafeez2021dpnct} ( A preliminary version has been published by IEEE International Conference on Communications (ICC) - Workshop on Communication, Computing, and Networking in Cyber-Physical Systems (IEEE CCN-CPS ), Montreal, Canada, June 2021, entitled “DPNCT: A Differential Private Noise Cancellation Model for Load Monitoring and Billing for Smart Meters”), whose core contribution removed the use of a trusted third party aggregator in DP scheme and enabling the calculation of highly accurate billing using a periodic noise cancellation technique at a low computational cost. In this extended work, we have modified DPNCT with split noise distribution over multiple smart meters, increasing the approach resilience against collusion attacks. We access E-DPNCT performance, by comparing it against a lightweight encryption based collusion resistant privacy solution, EPIC \cite{epic}. We chose EPIC as a comparison due to the lack of alternative DP collusion resistant approaches. We demonstrate through our analysis that E-DPNCT is collusion attack resistant, yields highly accurate results in billing and load monitoring with low computational cost.

\subsection{Contributions}
Our contributions are highlighted as follows:
\begin{itemize}
    \item We present E-DPNCT that is collusion attack resistant by splitting the noise over multiple master smart meters (MSMs). This, to the best of the authors knowledge, is the first data perturbation DP scheme for smart grids that does not make an assumption of trusted entities and that is resilient against collusion attacks.
    \item We assess the performance of our E-DPNCT against the state of the art encryption collusion resistant approach EPIC \cite{epic}. 
   {\item We compared our results in accuracy with DP based model "Differentially Private Demand Side Management for Incentivized Dynamic Pricing in Smart Grid (DRDP)" \cite{hassan2022}.}
    \item We experimented with multiple sensitivity values and study their impact on privacy and accuracy/utility of the data.
\end{itemize}
 \par

The rest of the paper is organized as follows. The related work is discussed in section \ref{litreview}. The threat model and Preliminary Knowledge are introduced in section \ref{threatmodelsection} and \ref{PrelimSection} respectively. After that proposed solution is introduced in section \ref{ProposedEDPNCTsection}. Finally, The paper is concluded with section \ref{Con} that includes summary of analysis on E-DPNCT and the future directions.
\begin{table}[h]
\footnotesize
\caption{Comparison of Techniques for Privacy Preserving using Differential Privacy in smart meters}
\begin{tabularx}{\linewidth}
 {|m{3em}|m{8em}|m{19em}|m{4em}|m{8em}|m{11.25em}|}
\hline
\textbf{Ref. No}& 
\textbf{\textit{Privacy Type}}& 
\textbf{\textit{Working Mechanism }}&
\textbf{\textit{Aggregator Type }}&
\textbf{\textit{Security Analysis}}&
\textbf{\textit{Limitation}}\\
\hline

\cite{T_Bicom}&
Differential Privacy&
Laplacian noise is added at trusted remote entity for effective and private bi directional communication between power grid and consumer.&
Trusted&
No privacy attacks analysis is available&
No privacy from trusted entity\\
\hline

\cite{Hassan2020,hassan2022}& 
Differential Privacy& 
Dual Differential Privacy (Laplacian Noise) with Dynamic pricing for fair billing using trusted third party& 
Trusted&
No privacy attacks analysis is available&
No privacy from aggregator, no analysis on the usability of differentially private data at grid level\\
\hline

\cite{Eibl2017}& 
Differential Privacy& 
Adding gamma distributed noise to each individual agent using infinite divisible Laplace distribution&
Not Trusted&
No privacy attacks analysis is available&
Privacy for aggregated information only\\
\hline

\cite{Hale2019}& 
Differential Privacy& 
Finding balance at individual level privacy with increased data points for decrease error in billing error&
Not Trusted &
No privacy attacks analysis is available &
Reduced accuracy in utility\\
\hline
\cite{Barbosa2016}& 
Differential Privacy& 
Differential privacy using Laplacian noise with filtering attack analysis to preserve appliance usage privacy& 
Not Trusted &
Filtering attack resistant &
Reduced accuracy in utility, No analysis on internal attacks \\
\hline
\cite{epic} & 
Hash MAC and Homomorphic Encryption& 
Aggregated load and bill calculation with privacy preservation of individual using multiple proxies and short term encrypted messages &
Not Trusted & 
Collusion attack resistant & 
High computational complexity and communication overhead \\ 
\hline
\cite{wang2020internalAttacks} & 
Encryption with Fill Function &
Lightweight internal attack resistant privacy preservation technique using homomorphic enryption with dynamic entry and exit for member smart meters&
Not Trusted & 
Collusion attack resistant & 
High computational complexity and communication overhead \\ 
\hline
\cite{dream}& 
DP with Encryption& 
Multiple exchange of encrypted messages with aggregator for DP masked data and cluster based analysis for privacy and utility analysis&
Not Trusted&
Collusion attack resistant&
Partial fault tolerance, 
increased utilization of bandwidth and
privacy for aggregated data only\\
\hline

\cite{Won2016}& 
DP with Encryption (Modular addition) & 
Differential privacy using Laplacian noise with current and future cipher text for fault tolerance with modular additive encryption&
Not Trusted&
No privacy attacks analysis is available&
Computationally Complex,
No privacy for individuals data profiles
\\
\hline

\cite{baza2021privacy} & 
Encryp. with Charge control devices &
Electric storage units used as proxies for encrypted charging request sharing at different time slots to preserve privacy &
Not Trusted & 
Collusion attack resistant& 
High computational complexity and Material Cost \\ 
\hline

\cite{Zhang} & 
Encryption &
$k$ partitioned encrypted data with spatial and temporal aggregation with $(k-1)$ distributed smart meters &
Not Trusted & 
Collusion attack resistant& 
High computational complexity and communication overhead \\ 
\hline

\cite{Mustafa} & 
MPC Encryption &
Multiple data aggregated techniques with a trade off between accuracy and privacy with secure multiparty computation algorithms &
Not Trusted & 
Collusion attack resistant& 
High computational complexity and Communication overhead \\ 
\hline

This paper & 
Differential Privacy &
Laplacian noise added at an instant is split into multiple parts and sent to master smart meters for aggregated noise cancellation at aggregator level along with self noise cancellation models accurate billing&
Not Trusted & 
Filtering Attack, Collusion attack & 
Communication overhead \\ 
\hline
\end{tabularx}
\label{tab1}
\end{table}

\section{Literature Review}\label{litreview}
The literature review is further divided into two parts. The first part provides an overview of existing privacy models for smart grids based on $(a)$ privacy technique i.e. DP, encryption, and hybrid; and $(b)$ aggregator type i.e. trusted and un-trusted third party aggregator, is presented.
In the second part, a discussion on the security analysis (i.e., their resistance against collusion attacks) of these privacy models is presented. 
 \par

\subsection{Privacy Models for Smart Grid}
Paverd \emph{et al.} \cite{T_Bicom} use a remote trusted entity to add Laplacian noise in smart meters data. This remote trusted entity is responsible for bi-directional communication between the power grid and the smart meter for an effective Demand Response (DR) mechanism. 
{Dynamic billing to reward correct behaviour and enforce demand response model is proposed by the authors from \cite{Hassan2020,hassan2022}. } They provide DP at aggregator level where a trusted aggregator collects original data and Laplacian noise is generated and added to the original data. Dynamic bills are calculated using original data and only the customers responsible for peak load are charged with peak factor price to ensure fair billing. However, a trusted entity is required in both \cite{T_Bicom} and \cite{Hassan2020} models to mask the original data and follow the demand response protocol honestly.  {Z. Liu \emph{et al.} \cite{com3} uses zero knowledge proof and a trusted authority which is  responsible for registering users and public and private key management. }\par 
The solutions with non trusted third party including \cite{Eibl2017} and \cite{Hale2019} used infinite divisibility of Laplacian distribution and point-wise sensitivity to generate and add noise at the smart meter level. The contributions in \cite{Eibl2017} and \cite{Hale2019} were limited in that DP was discussed only within the context of aggregated data for load monitoring, and did not detail the subsequent impact of the noise and accuracy of billing to the end user, nor was a security analysis of the approach presented.The BDP model presented in \cite{Barbosa2016} also uses DP for the preservation of appliance usage privacy. BDP privacy preservation model is focused on masking appliance usage of a households by choosing sensitivity as the maximum wattage of the heaviest electrical appliance. {Ren  \emph{et al.} \cite{com4} uses a novel measurement based perturbation for accuracy in bills. However, the paper did not discuss the impact of noise addition on load monitoring in the experiments.  }  \par 

Similar to \cite{Eibl2017,Hale2019} and \cite{Barbosa2016}, EPIC by Alsharif \emph{et al.} \cite{epic} and Wang \emph{et al.} \cite{wang2020internalAttacks} use a non trusted aggregator. They differ in their privacy mechanism as they only use compute intensive encryption based on a key exchange mechanism which has a greater communication and computation overhead as compared to pure DP based solutions. Similar to \cite{Eibl2017,Hale2019} and \cite{Barbosa2016}, EPIC by Alsharif \emph{et al.} \cite{epic} L. Wu \emph{et al.} \cite{com2}, and Wang \emph{et al.} \cite{wang2020internalAttacks} use a non trusted aggregator. They differ in their privacy mechanism as they only use compute intensive encryption based on a key exchange mechanism which has a greater communication and computation overhead as compared to pure DP based solutions. { X.-Y. Zhang \emph{et al.} \cite{com1} proposed service model that trains the neural network models locally, and only model parameters are shared with the central server instead of sending private energy data to the cloud server. The goal of the paper is however forecasting of energy demand and federated learning model predicts future energy demand based on multiple features including current demand, weather etc. } \par 

Acs \emph{et al.}\cite{dream} and Won \emph{et al.}\cite{Won2016} also use a non trusted aggregator in their approach. They differ though in that as they proposed a hybrid approach, using encryption in addition to differential private noise between the smart meters and aggregator, to mask the data. However, these solutions are computationally complex and consume extra bandwidth in the network to send $ciphertexts$ information. The authors from \cite{baza2021privacy} make use of encryption and scheduling of charging batteries as a privacy mechanism without a trusted third party. This solution requires extra material cost for installing and maintaining energy storage devices such as batteries.  \par
DPNCT \cite{hafeez2021dpnct} used DP with noise cancellation without a trusted third party for accurate and private load monitoring and billing. This model is put under the test of collusion attacks in this paper and it proves to be vulnerable to collusion attacks in case of malicious smart meters. Enhancing the attack resistance of the noise cancellation model, E-DPNCT makes use of split noise cancellation and variable privacy selection for the electricity consumers which makes it resistant to collusion attacks.

\begin{figure*}[t]
\begin{subfigure}[t]{.5\textwidth}
\centering
   \includegraphics[width=0.9\linewidth]{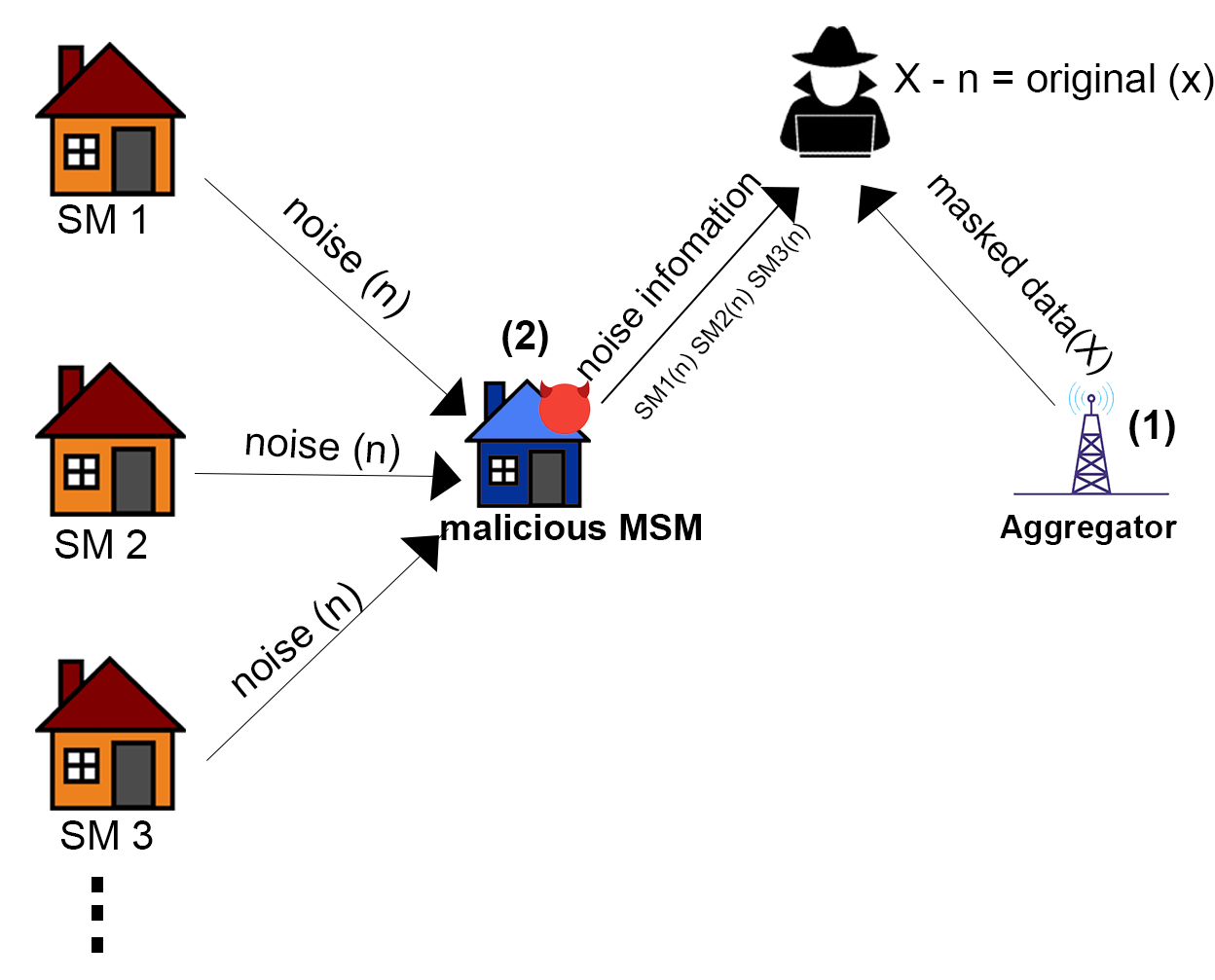}
   
   \caption{DPNCT}
   \label{fig:CAScene1} 
\end{subfigure}
\begin{subfigure}[t]{.5\textwidth}
   \includegraphics[width=0.9\linewidth,]{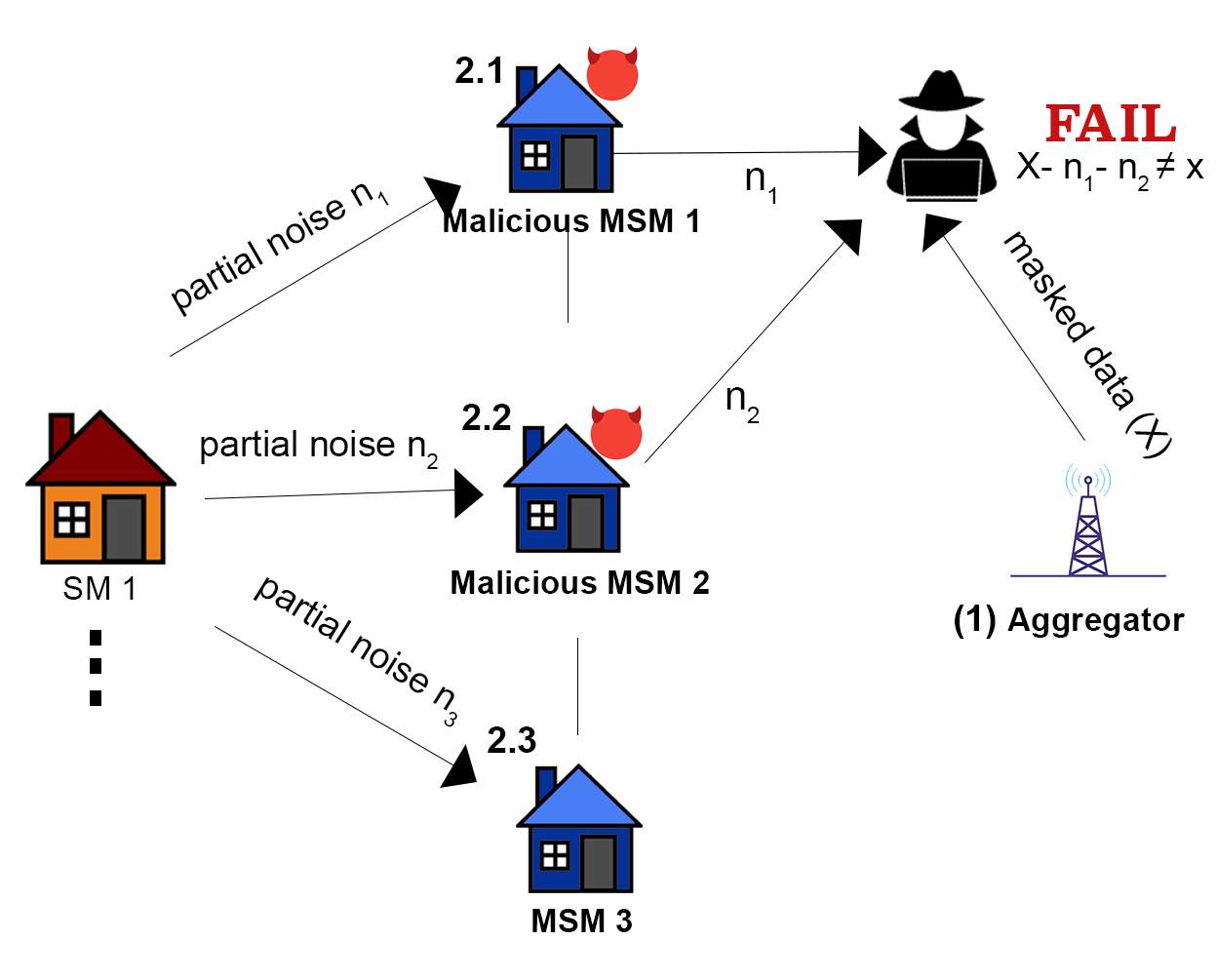}
   \caption{E-DPNCT with $3$ MSMs.}
   \label{fig:CAScene2}
\end{subfigure}
\caption{Collusion attack scenarios}
\end{figure*}
\subsection{Security Analysis of Privacy Models}
Table \ref{tab1}, presents a summary of privacy mechanism in smart grid, highlighting a brief overview of its operation and the aggregator type along with a critical analysis of the main limitations of the approach and available security analysis. The following paragraphs also highlight key DP and encryption based privacy models where a security analysis has been performed on them to assess their resilience against data privacy or integrity security attacks.


\subsubsection{DP or Hybrid Privacy Models}
The BDP model \cite{Barbosa2016} proposed differential privacy with trusted third party aggregator. Barbosa \emph{et al.} simulated filtering attack on the protected data of $200$ households. Their analysis shows that high level of differential privacy protects the differentially private data against filtering privacy attacks.  \par

As for collusion attack resistance privacy models, \cite{dream} is $(n-1)$ collusion resistant against data reconstruction attack that utilises DP. However, they used a hybrid approach, as they used a data perturbation differential privacy technique to add noise first and then used sharing secret key for data encryption and decryption at the aggregator end. Other collusion attack resistant privacy models \cite{Zhang,Mustafa,wang2020internalAttacks,baza2021privacy,epic} utilise only encryption. \par 
\subsubsection{Encryption Privacy Models}
Zha \emph{et al.} \cite{Zhang} proposed an encryption based privacy model that is resilient to internal attacks where aggregators as well smart meters are assumed to be malicious. Mustafa \emph{et al.} \cite{Mustafa} used multiparty computation algorithms (MPC) for privately aggregating electricity consumption data. Their model is collusion resistant for up to two third of malicious parties using a verifiable secret sharing technique. The privacy model proposed by \cite{wang2020internalAttacks} used session keys and a fill function in a way that the aggregated mask becomes zero at the non trusted aggregator. In order to be protected from internal collusion attacks they used encryption, and their mathematical model achieves reliable privacy protection against collusion attack. {L. Wu \emph{et al.} \cite{com2} uses HTV-PRE, a homomorphic threshold proxy re-encryption scheme with re-encryption verifiability for privacy preservation in smart grids. }Baza \emph{et al.} \cite{baza2021privacy} is resistant to collusion attack by the means of Partial Blind Signature (PBS) during the acquisition of anonymous tokens and the one time generated identity is not link-able to the charging unit. The privacy preserving model introduced by them is for charging coordination of batteries only. The authors from \cite{epic} proposed EPIC and introduced the idea of $proxies$  where each smart meter selects a number of $proxies$ and sends them small chunks of pairwise secret masks. They analysed the impact of  collusion attack on EPIC using hyperbolic probability model. All the collusion resistant privacy models in smart meters uses some form of encryption to protect them which increases the computational overhead of the solution.  \par
For collusion resistance, E-DPNCT is compared with EPIC \cite{epic} as the system model used by them is similar to our model where both models share information with randomly selected MSMs. In E-DPNCT, instead of sharing partial encrypted electricity consumption with MSMs, each smart meter shares DP noise with the master smart meters (MSMs). Considering the E-DPNCT only uses DP which is not compute intensive, is a better solution in terms of efficiency, accuracy and security. 

 \section{Threat Model and Attacks} \label{threatmodelsection}
As mentioned previously, a collusion attack is an attack where the adversary conspires with entities of the smart grid in order to retrieve the original time series data of users' energy consumption which poses a threat to the privacy of electricity consumers \cite{PurturbAttackRECon}. \par
In the attack scenario considered, the goal of the adversary is to find real time energy usage data of individual consumers, to analyse the pattern and infer sensitive information from it. In the threat model, the aggregator is assumed to have full access to the masked electricity consumption profiles data of consumers and are also assumed to be honest, but not trusted entities hence, they can try to infer information from masked data but they will not alter it. The smart meters choose masters among other smart meters for privately sharing masking information with the aggregator. The smart meters can be malicious and hence, can share masking information with an adversary if selected as a Master Smart Meter (MSM). The aggregator along with colluding MSMs may try to launch a collusion attack by sharing their private noise with an adversary. \par

\subsection{DP, DPNCT and Collusion Attack Resilience}
  \par
In DP privacy models, at every instant $t$ each smart meter generates and adds DP noise into its energy consumption data before sending it to the aggregator for load monitoring and billing. A noise cancellation technique is adopted from DPNCT \cite{hafeez2021dpnct} where each smart meter sends the added noise to a randomly selected master smart meter. The aggregated noise from the MSMs is then sent to the aggregator to calculate total load in an area for load monitoring. \par

The privacy model adopted in DPNCT is vulnerable to collusion attacks, if the aggregator and MSMs collude to compute the original reading of the individual smart meters. This is further demonstrated in Fig. \ref{fig:CAScene1}, where an adversary can get masked data from the aggregator $(1)$ and collude with a malicious MSM to get individual noise information $(2)$ added by each smart meter at an instant $t$. The added noise can be subtracted from individual masked profile to get the original energy consumption data. 

\begin{figure*}[t]
\centerline{\includegraphics[scale=0.09]{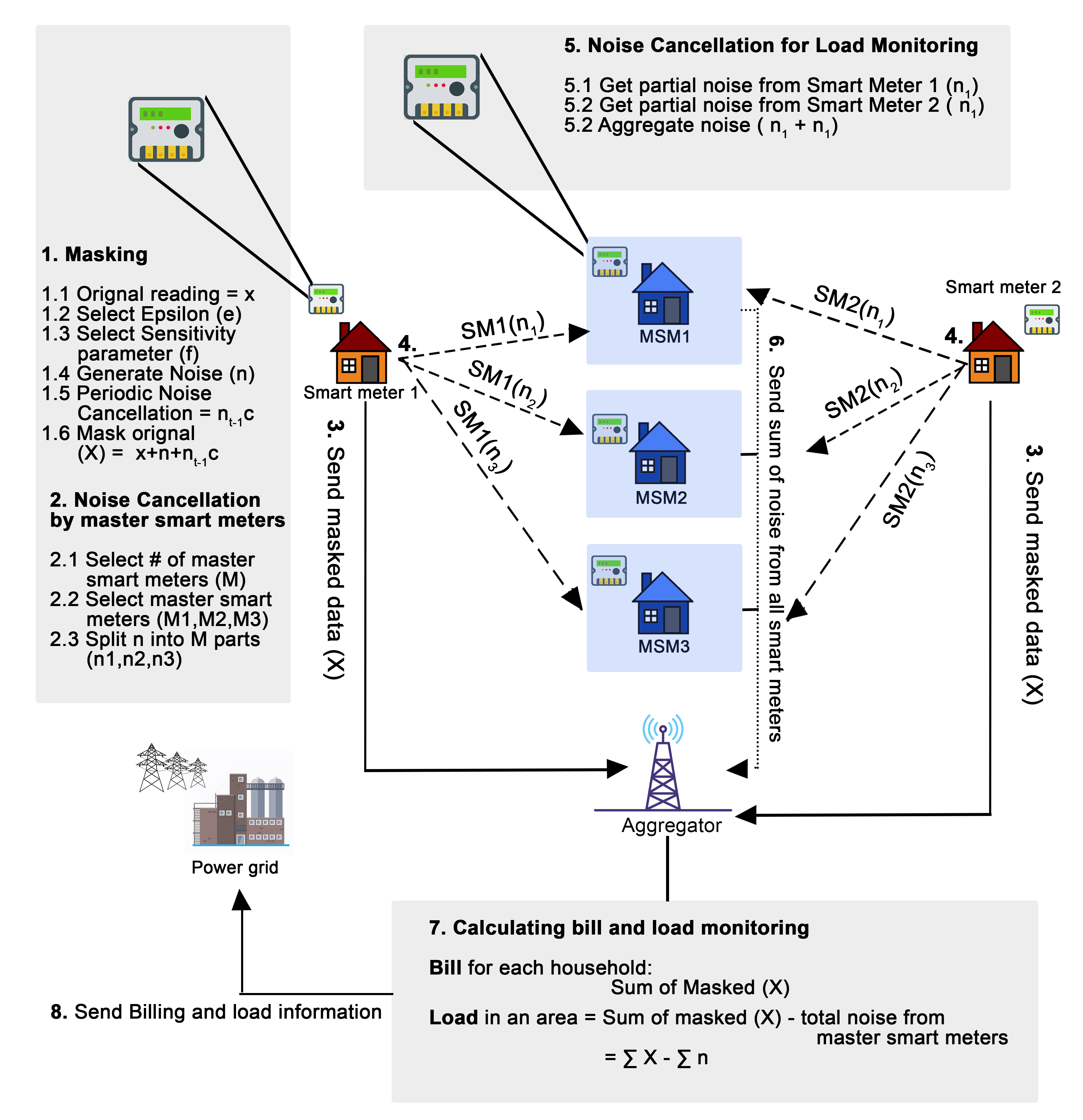}}
\caption{System model of split noise E-DPNCT.}
\label{sm}
\end{figure*}

\begin{figure}[t]
\centerline{\includegraphics[width=0.5\linewidth]{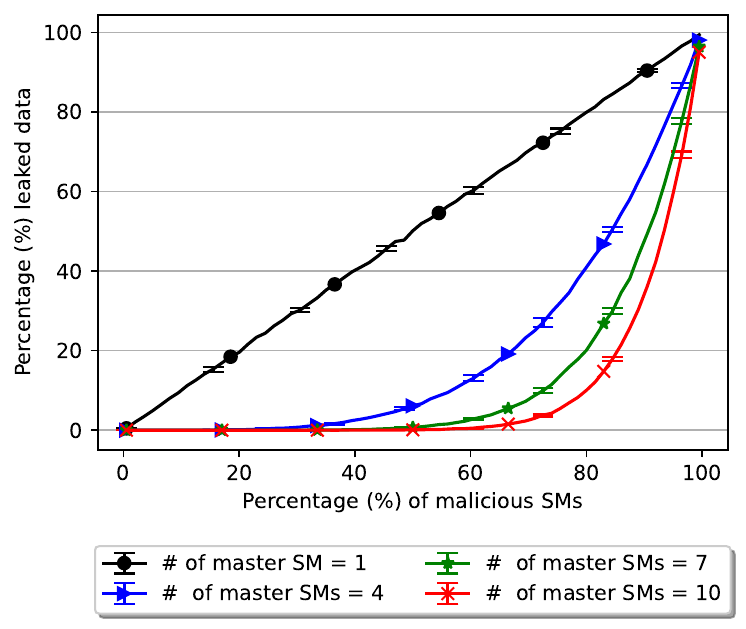}}
\caption{Collusion attack on E-DPNCT with multiple master smart meters.}
\label{modifiedCA200}
\end{figure}


\begin{figure*}[t]
\begin{subfigure}[t]{.3\textwidth}
  \centering
\includegraphics[width=1\linewidth]{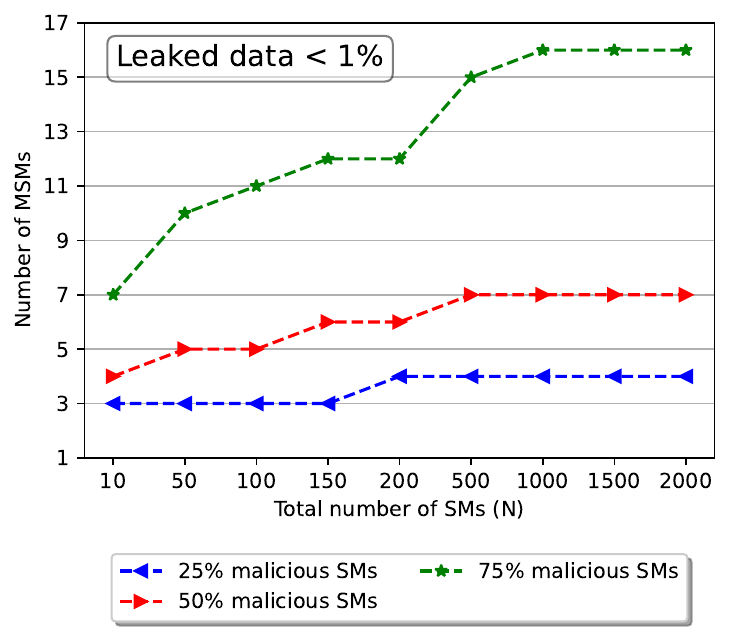}
\caption{Required number of MSMs for\\ less than 1$\%$ data leak}
\label{prop1}
\end{subfigure}
\begin{subfigure}[t]{.3\textwidth}
  \centering
  \includegraphics[width=1\linewidth]{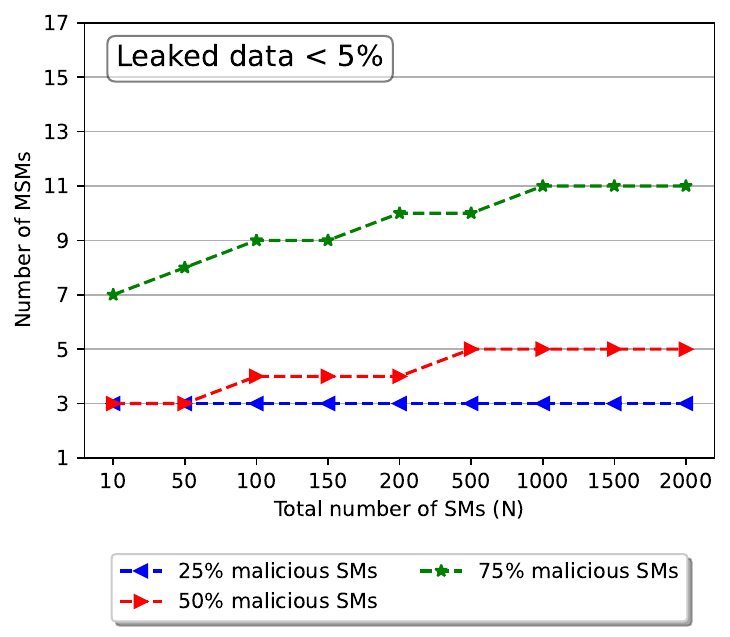}
  \caption{Required number of MSMs for\\ less than 5$\%$ data leak} 
\label{prop5}
  \end{subfigure}
  \begin{subfigure}[t]{.3\textwidth}
  \centering
  \includegraphics[width=1\linewidth]{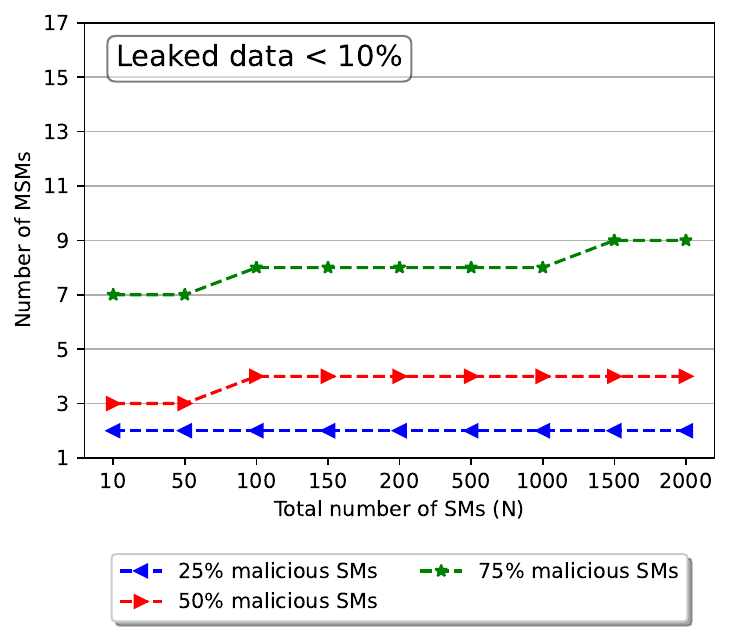}
  \caption{Required number of MSMs for\\ less than 10$\%$ data leak} 
\label{prop10}
  \end{subfigure}
  \caption{Rate of increase in required number of MSMs for successful collusion attack resistance}
\end{figure*}

\begin{figure}[t]
\centerline{\includegraphics[width=0.5\linewidth]{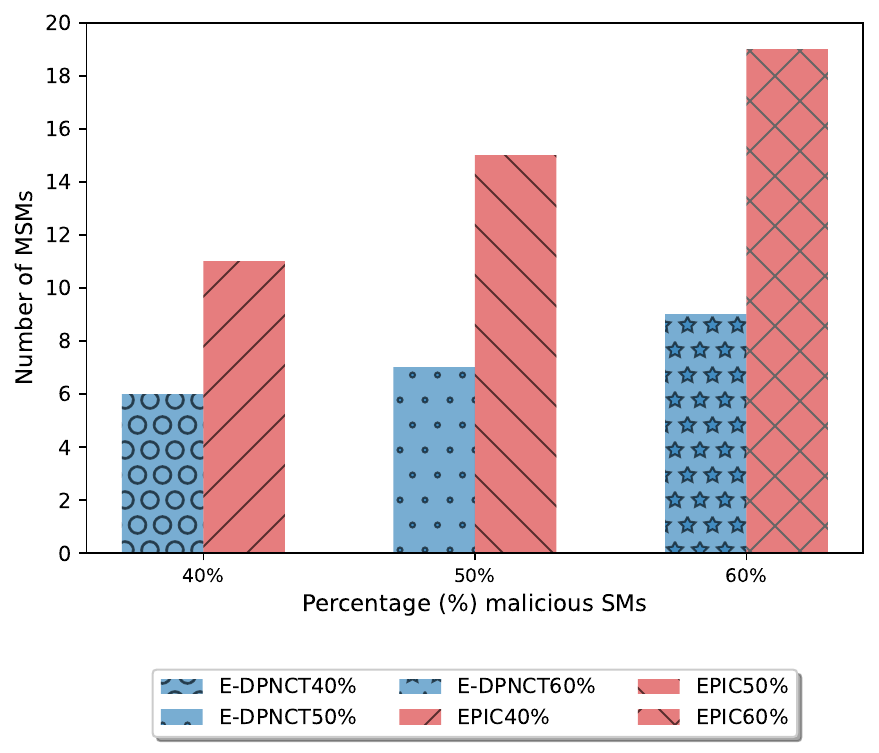}}
\caption{Comparison of collusion attack resistance between split noise E-DPNCT and EPIC \cite{epic}. }
\label{DPNCvsEPIC}
\end{figure}

\begin{algorithm}[h]
\small
\SetAlgoLined
\SetKwFunction{FnENC}{Function E-DPNCT()}
\SetKwFunction{FnSplit}{Function RandomlySplitNoise()}
\FnENC \;
\Begin {
\KwIn{$x_t,ID,\Delta t,masterIDs_t,totalBill,surchargeUnits$}
$N_{t-1}$ = $N_{t}$\;
$N_{t}$ = 0\;
 \While{Time Period $\Delta t$}{
 Select $\Delta f$ from ${max,1/2 max, avg, 1/2 avg}$\;
 Calculate $\lambda$ using $\Delta f$ and $\epsilon$\;
 $n_t$ = G(N,$\lambda$) - G'(N,$\lambda$)\;
 Push($n_t$) to $N_{t}$\;
 $Split_N{t}$ = RandomlySplitNoise($N_t,m$)\; 
 $nc_{t-1}$ = Pop($N_{t-1}$)\;
 $X_t$ = $x_t$ + $n_t$ - $nc_{t-1}$\;
 Send $X_t$ to aggregator \;
 \eIf{$ID$ in $masterIDs_t$ }{
 \For{all $k$ smart meters in group}{
 get noise $n_{k,t}$ from $kth$ smart meter\;
 }
 Report aggregated group noise $ \sum_k n_{k,t}$ to aggregator
 }{
  \For{all $m$ master smart meters in $masterIDs_t$}{
    $SplitNoise_{t}$ = Pop($Split_N{t}$)\;
    Send $SplitNoise_{t}$ to MSM with $masterID_{t,m}$\;
 }
 
 }
 }
}
\FnSplit \;
\Begin {
number of MSM = $m $\;
partial Noise list = $dirichlet_distribution(m,totalNoise_t) $\;
$return$ 
$ partial Noise list$\;
}
\caption{Enhanced DPNCT}
 \label{alg:DPNC3}
\end{algorithm}

\begin{algorithm}[h]
\small
\SetAlgoLined
\SetKwFunction{FnB}{Function BillCalculation()}
\SetKwFunction{FnL}{Function AggregatedLoadCalculation()}

\FnL \;
\Begin {
\KwIn{ $mIDs$} 

 \While{Billing Period $T$}{
 \For{all smart meters $i$ in $N$}
 {
  $X_i$ = getMaskedData($i$)\;
 
  }
  \For{$masterID$ in $mIDs$}{
  $N_K$ = getNoiseData($masterID$)\;
  }
  $TotalLoad_{t}$ = $\sum_{i=1}^{N} X_i - \sum_{i=1}^{K} N_K$ \;
 }
 }
 \FnB \;
\Begin {
\KwIn{ $maxUnits, SurchargePrice, UnitPrice$,$Er_{T-1}$}
 \For{ all smart meters $i$ in $N$}{
 \eIf{$\sum_{i}^{T} X_i \geq maxAllaowedUnits$}{
 
 $surchargeUnits = \sum X_i - maxAllaowedUnits$\;
  
 $BaseBill$ = $maxAllaowedUnits * UnitPrice $\;
 $SurchargeBill$ = $surchargeUnits *SurchargePrice$\;
 $TotalBill_i$ =$BaseBill + SurchargeBill - Er_{T-1}$\;
 Notify $TotalBill_i$ and $surchargeUnits$ to smart meter $i$\;
 
 }{
 $TotalBill_i$ = $\sum X_i * UnitPrice$\;
 Notify $TotalBill_i$ to smart meter $i$ \;
 }
 }
 }

 \caption{Calculation of Bill and Aggregated Load at Aggregator}
 \label{alg:DPNC1}
\end{algorithm}

\begin{figure*}[t]
\begin{subfigure}[t]{.5\textwidth}
  \centering
   \includegraphics[height = 220pt,width=1\linewidth]{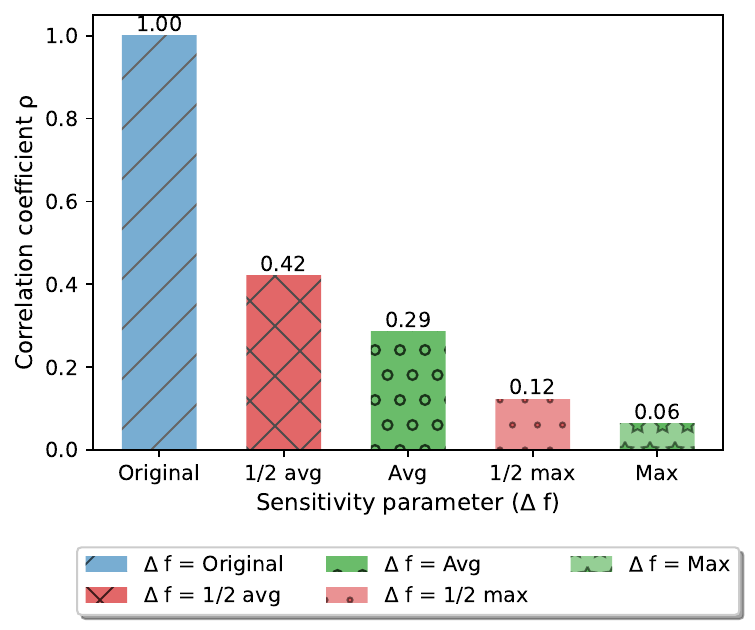}
    \caption{Impact of different sensitivity values on Privacy of data }
 \label{fig:privacyvsSensitivity}
  
\end{subfigure}
\begin{subfigure}[t]{.5\textwidth}
  \centering
  \includegraphics[ height = 220pt, width=1\linewidth]{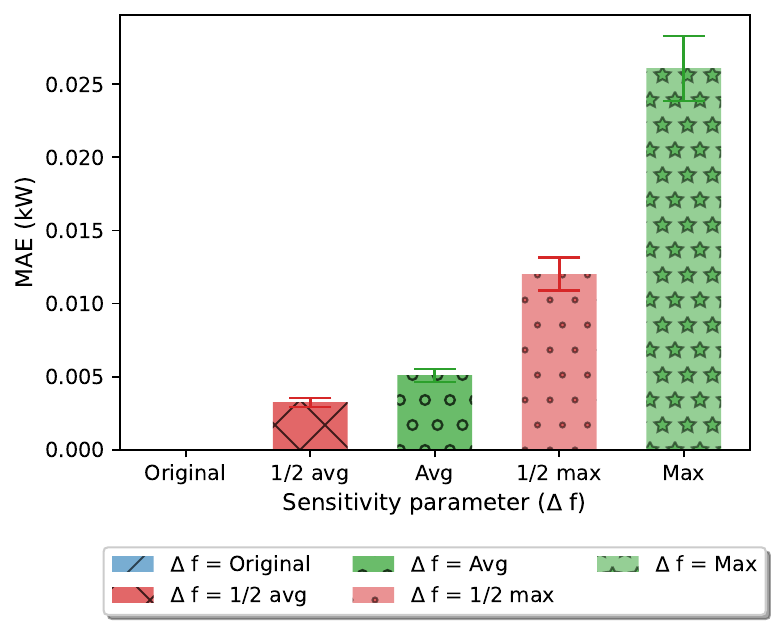}
    \caption{Impact of different sensitivity values on accuracy in billing}
  \label{fig:UtilityvsSensitivity}
\end{subfigure}
\label{}
\caption{Impact of sensitivity on privacy and accuracy. }
\end{figure*}
\section{Preliminary Knowledge}\label{PrelimSection}
In this section preliminary information on differential privacy is discussed which is used construct to construct E-DPNCT. 
\subsection{Differential Privacy}
The probabilistic model of Dwork \emph{et al.} \cite{dwork2014algorithmic} states that the DP protected data ensures privacy for a mechanism $M$ for any two neighbouring data sets $D1$ and $D2$ that differ in one record and for all the possible outcomes $S \subseteq Range (M)$, if the below equation \ref{Eq:1} is satisfied  \cite{dwork2014algorithmic}: 
\begin{equation}\label{Eq:1}
Pr(M(D1) \in  S) \leq e^\epsilon * Pr (M(D2) \in S)
\end{equation}
This ensures that if a query function $f$ runs on the neighbouring data sets $D1$ and $D2$ then the outputs are indistinguishable by the differential privacy mechanism $M$ where $\epsilon$ is the privacy budget.
\subsubsection{Laplace Mechanism}

The Laplace mechanism ensures differential privacy by outputting a query as $f(x)+n$ where $n$ is noise drawn from Laplace distribution $ f(x,\lambda) = 1/2 (e^{|x|/\lambda})$, where $\lambda = \Delta f/\epsilon$ and $\Delta f  $ is sensitivity of query over data $D$.  \par

In smart grids where aggregators are considered un-trusted entity each smart meter mask its own reading before sending to aggregator using Infinite divisibility principal of Laplace distribution. According to the infinite divisibility property of the Laplacian noise, if the sampling of a random variable is done from the probability distribution function of Laplace distribution then for $N \geq 1$, the distribution is infinite \cite{hafeez2021dpnct}:

\begin{equation}\label{Eq:3}
Lap(\lambda) = \sum_{i=1}^{N} (G(N,\lambda) - G'(N,\lambda))
\end{equation}

In this equation \ref{Eq:3} , $G$ and $G'$ represent identically distributed and independent gamma density functions having same parameters, $N$ represents the number of smart meters within the network and the selection of $\lambda$ is based on and point-wise sensitivity. Equation \ref{Eq:3} states that when using gamma density function, the aggregated noise of all the smart meters at the network level will be equal to $Lap(\lambda)$ at time $t$.

\subsubsection{Sensitivity}
By definition, sensitivity refers to the maximum difference in the output of two neighbouring data sets for a function $f$, defined further in Equation \ref{Eq:2} \cite{dwork2014algorithmic}.
\begin{equation}\label{Eq:2}  
\Delta f = max {f(x)_{D1,D2} | f(D1) - f(D2) |  }
\end{equation}
Sensitivity $(\Delta f)$ is dependent on the function $f$ and the type of data. In smart grids the function $f$ is total electricity consumption of an area at an instant $t$ which is load monitoring and total energy consumption by a household in billing period $T$. Most commonly the sensitivity is selected as the maximum amount a household can consume in an area. 
\subsubsection{Sequential Composition}
If $M_1(D)$ satisfies $\epsilon 1$ differential privacy and $M_2(D)$ satisfies $\epsilon 2$ differential privacy then the combined mechanism that releases both outputs satisfies $\epsilon 1 + \epsilon 2$. 
\subsubsection{Parallel Composition}
If $M(D)$ satisfies $\epsilon$-DP and $D_1$ and $D_2$ are two disjoint subsets of $D$ such that $D_1 \cap D_2 = D$ then the mechanism which releases all of the results of disjoints sets satisfies $\epsilon$-differential privacy.
{\subsubsection{$\epsilon$-DP Guarantee Theorem}
Differential private metering reporting in E-DPNCT satisfy $\epsilon$-DP guarantee. The proof is as follows:\\
Let us consider {$F,F^\prime \in R^{|X|}$} in a way such that $||F - F^\prime||_{1} \leq 1$ and $F = {x_1,x_2,x_3, ..., x_n}$. Let $M$ be a function such that $ M: R^{|X|} \rightarrow  N^{k}$. $F$ and $F^\prime$ can be represented by their probability density functions linked with Laplace distribution as $pF_1$ and $pF_2$. According to \cite{dwork2014algorithmic}, these probability distributed functions can be compared as follows: \\
\begin{equation}
\frac{p_{F_{n}} \left[F = \lbrace x_{1}, x_{2},.., x_{n}\rbrace \right]}{p_{F_{n}^\prime }\left[F^\prime = \lbrace x_{1}, x_{2},.., x_{n}\rbrace \right]} = 
\end{equation} 
\begin{equation}
    \prod _{j=1}^{k} \frac{\exp \left(- \frac{\varepsilon |M(F_{n})_{j} - x_{j}|}{\Delta f}\right)}{\exp \left(- \frac{\varepsilon |M(F_{n}^\prime)_{j} - x_{j}|}{\Delta f}\right)} 
\end{equation}

\begin{equation}
    = \prod _{j=1}^{k} \exp \left(\frac{\varepsilon (|M(F_{n}^\prime)_{j} - x_{j}| - |M(F_{n})_{j} - x_{j}|)}{\Delta f}\right)
\end{equation}
\begin{equation}
    \leq \prod _{j=1}^{k} \exp \left(\frac{\varepsilon (|M(F_{n})_{j} - |M(F_{n}^\prime)_{j} |)}{\Delta f}\right)
\end{equation}
\begin{equation}
    = \exp \left(\frac{\varepsilon (||M(F_{n}) - |M(F_{n}^\prime)||)}{\Delta f}\right)
\end{equation}
\begin{equation}
    \leq \exp (\varepsilon)
\end{equation}
}
\section{Methods} 
\label{ProposedEDPNCTsection}
In this section our proposed collusion resistant E-DPNCT privacy model is introduced with reference to Fig. \ref{sm} and Algorithm \ref{alg:DPNC3}. Its performance against collusion attacks is assessed and compared to an encryption based approach EPIC \cite{epic}. In addition, we analyse E-DPNCT privacy of individual consumers and accuracy in providing billing an load monitoring.
\subsection{E-DPNCT Operation}
A step by step split noise collusion resistant E-DPNCT model is introduced in Fig. \ref{sm}. In E-DPNCT, as shown in Step $1$, each smart meter will firstly select privacy parameters to mask the original data by generating DP noise using Laplace distribution and adding this noise to the original metered data before sending it to the aggregator. 
In case of E-DPNCT, the query function $f$ can be bill calculation over a period of time $T$ or load monitoring for $N$ number of households in an area. If DP noise is added to each individual smart meters reading then according to the above equation \ref{Eq:1} it ensures $\epsilon$-DP protection. \par

As for generating noise through Laplace Mechanism, a random variable is generated from probability density function of Laplace distribution. 
In E-DPNCT (Fig. \ref{sm}, Step $1.2$), the privacy parameter $\epsilon$  is controlled by the user through a defined range from $0 - 1$. The smaller value of $\epsilon$ ensures more privacy, this however, comes at the cost of errors in accuracy of billing and load monitoring. 

Laplace mechanism employed in E-DPNCT relies on another privacy parameter which is sensitivity $(\Delta f)$, the setting of which depends on the type of data and query. In case of E-DPNCT, the two functions are billing and load monitoring which relies on sum of all measurements so the sensitivity is the maximum difference a single measurement can make on billing and load monitoring. It is calculated in multiple different ways, the most common of which, is the maximum measurement among all the smart meters. \par 
In E-DPNCT (Fig. \ref{sm}, Step $1.3$), when adding the noise, each smart meter can choose its level of sensitivity as a privacy parameter ensuring a personalised level of privacy and accuracy trade off. Since different households have different amount of energy consumption and require different level of privacy, a robust and personalised sensitivity level ensures a more personalised level of privacy. Considering $N$ households have different energy consumption ${x_1,x_2,x_3,...,x_N}$ at an instant $t$, the sensitivity parameter ($\Delta$ f) can be chosen as a maximum value from ${x_1,x_2,x_3,...,x_N}$ or a mean of ${x_1,x_2,x_3,...,x_N}$ according to the requirements of privacy and accuracy. In E-DPNCT, the following values are experimented with sensitivity parameter: 
\begin{itemize}
    \item $\Delta f$ = max ${x_1,x_2,x_3,...,x_N}$ 
    \item $\Delta f$ = $\frac{max}{2} $ ${x_1,x_2,x_3,...,x_N}$ 
    \item $\Delta f$ = average ${x_1,x_2,x_3,...,x_N}$  = $\frac{\sum{x_1,x_2,x_3,...,x_N}}{N}$
    \item $\Delta f$ = $\frac{average}{2} $   
\end{itemize}


{The mechanism is detailed in Algorithm \ref{alg:DPNC3} function $E-DPNCT()$ where each smart meter selects a sensitivity parameter $\Delta f$ and generate noise $n_t$ at instant $t$. Noise $nc_t$ from previous time period $\Delta t$ is subtracted and $n_t$ is added to original reading $x_t$. $n_t$ is split into $m$ parts using the function $RandomlySplitNoise$ and send to $m$ selected master smart meters. $n_t$ is added to a list $N_t$ to keep track of total noise added in a time period  $\Delta t$ }. Further discussion on impact of sensitivity over privacy and accuracy is discussed later in section \ref{UitlityA}.

\begin{equation}\label{Eq:4}
\Delta f = \{ max_{i,t} |x_{i,t}|, avg_{i,t} |x_{i,t}|,\frac{max_{i,t} |x_{i,t}}{2}|, \frac{avg_{i,t} |x_{i,t}|}{2} \}
\end{equation}

Using both $\epsilon$ and $sensitivity$ the noise ($n$) is generated (Fig. \ref{sm}, Step 1.4). Next, noise added in previous time period is collected as $n_{t-1}c$ (Fig. \ref{sm}, Step 1.5). Noise $n$ is added and $n_{t-1}c$ is cancelled in original data $x$ to mask it (Fig. \ref{sm}, Step 1.6).  


In Step $2$, it can be seen in Fig. \ref{sm} that each smart meter selects $m$ number of MSMs. For attack resistance against colluding smart meters, each smart meter splits its noise into $m$ parts. The smart meter then sends the masked data $X$ to the aggregator (Fig. \ref{sm}, step $3$) and as Step $4$ in Fig. \ref{sm}, it then sends the partial noise to the selected $m$ MSMs at an instant $t$, which is further explained in Algorithm \ref{alg:DPNC3} function $RamdomlySplitNoise()$. 
In Step $5$ (Fig. \ref{sm}), each MSM aggregates the partial noise received from each smart meter and sends it to the aggregator shown in Step $6$ (Fig. \ref{sm}). The aggregator then aggregates masked data $X$ received from each smart meter in the area at instant $t$ and total noise $\sum n$ received from each MSM for the instant $t$. Aggregated noise data is subtracted from aggregated masked data to get total noise as shown in Step $7$ (Fig. \ref{sm}). The total bill is calculated by aggregating masked data $X_i$ per household for a billing period $T$. This mechanism is explained in Algorithm \ref{alg:DPNC1}. The mechanism of periodic self noise cancellation for billing is adopted from DPNCT \cite{hafeez2021dpnct} where each smart meter periodically cancels the noise added in the previous time period. The mechanism ensures accuracy in bills and is briefly explained in Algorithm \ref{alg:DPNC3}. {Function $AggregatedLoadCalculation$ takes masked readings $X_t$  from all smart meters in an area at an instant $t$ Aggregator collects aggregated noise $N_k$ from each master smart meter. The aggregated noise is then subtracted from masked data to get total load at an instant $t$. Similarly, using function $BillCalculation$ bills are calculated by aggregating masked reading $X_i$ by a smart meter $i$ for a billing period. If total units consumed are more than allowed units, the excess units are charged at surcharge unit price. }Billing and load monitoring analytic reports are then sent to the power grid for demand response policies. 

 \begin{figure}[t]
\centerline{\includegraphics[width=0.5\linewidth]{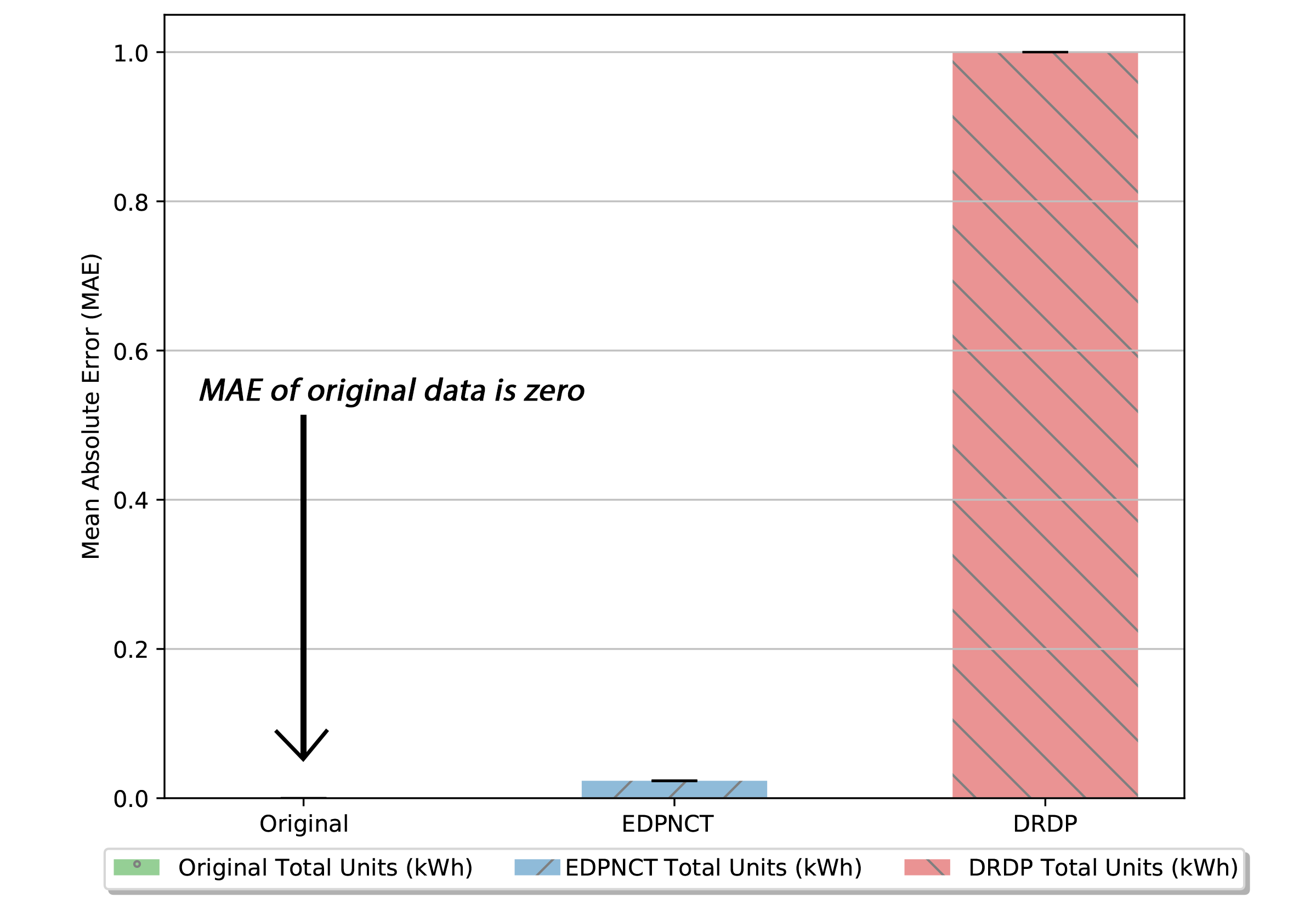}}
\caption{Comparison of mean absolute error in billing with DRDP \cite{hassan2022}. }
\label{fig:maeDPNCT}
\end{figure}

\begin{figure*}[t]
\begin{subfigure}[t]{.5\textwidth}
  \centering
   \includegraphics[width=1\linewidth]{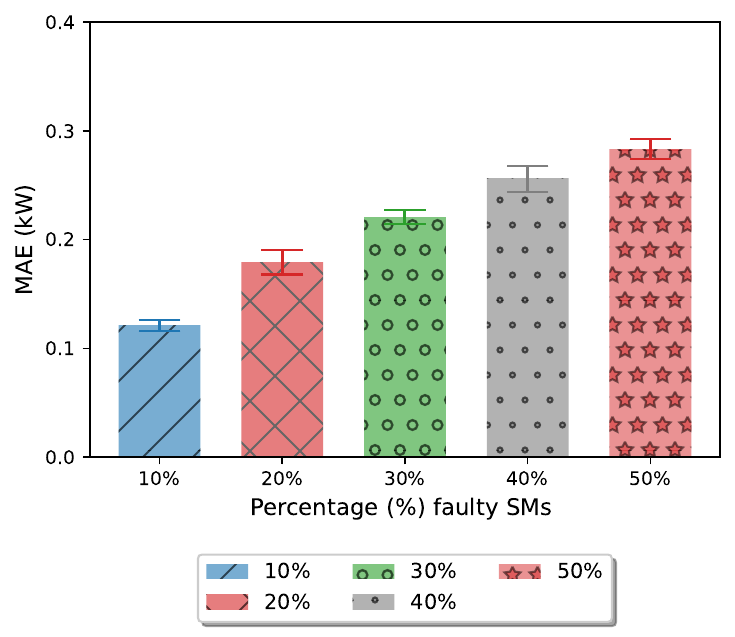}
    \caption{Comparison of mean absolute error in load monitoring for multiple levels of faulty smart meters.}
 \label{fig:maelm} 
\end{subfigure}
\begin{subfigure}[t]{.5\textwidth}
  \centering
\includegraphics[height = 220pt,width=1\linewidth]{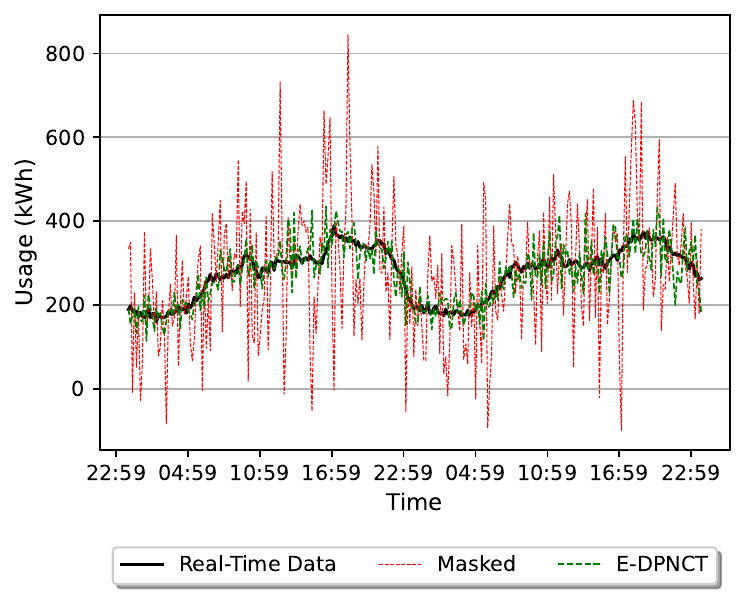}
    \caption{Load monitoring with $10\%$ faulty smart meters with E-DPNCT}
  \label{fig:lm}
\end{subfigure}
 

\label{fig:mae}
\caption{Load monitoring in E-DPNCT.}
\end{figure*}
\section{Experiments and Results}
In this section we discuss our experiments and their results with respect to resistance against collusion attack, level of privacy preservation and accuracy in utility functions i.e., billing and load monitoring. 
\subsection{Collusion Attack Resistance}
To assess the resistance of our split noise E-DPNCT against collusion attacks, a collusion attack is launched with multiple MSMs. As shown in Fig. \ref{fig:CAScene2}, a collusion attack on the E-DPNCT model is successful only if all the $m$ MSMs are malicious and colluding with the aggregator. In this scenario, the attacker needs to collude with all three of the MSMs to get complete noise information $(2.1,2.2,2.3)$ from Fig. \ref{fig:CAScene2} and the aggregator to get masked profile of the individual smart meters in order to compute original data. \par
At instant $t$, $m$ MSMs are randomly selected from the total smart meters $N$ in an area such that $m \subset N$. The probability that the selected MSM would be a colluding smart meter increases with the increase in total number of malicious smart meters in the group. Fig. \ref{modifiedCA200} shows the 
rate of success of a collusion attack with increasing percentage of malicious smart meters on x-axis and percentage leaked data on y-axis. As shown in this figure, with $m=4$ MSMs (blue line), the percentage leaked data in one month is less than $1\%$ when $50$ out of $200$ smart meters are malicious which is significantly better where only $1$ MSM (black line) is used. As the number of MSMs  $m$ increases, the resilience against collusion attacks also increases i.e. with $13$ MSMs, $135$ out of $200$ smart meters would need to be malicious before the percentage of leaked data increases over $1\%$.\par

A key question emerging from this research is the number of MSMs ($m$) required in the E-DPNCT model for successful resistance against collusion attacks. In order to answer this question, different scenarios were simulated varying the number of MSMs (y axis) as shown in Fig. \ref{prop1}. It can be demonstrated that with a large network size of $2000$ smart meters (x-axis) and $50\%$ malicious smart meters, a very small number of MSMs i.e. $m=7$ (red line), is required for a successful resistance against collusion attacks. Resistance against collusion attacks can be considered successful when only less than $1\%$ data is leaked. This definition can be relaxed to $5\%$ data leak and $10\%$ data leak. In Fig. \ref{prop1}, \ref{prop5} and \ref{prop10}, the required number of MSMs for successful resistance against collusion attacks are compared for $1\%$ data leak, $5\%$ data leak and $10\%$ data leak respectively. It can be seen from the results that as we relax boundaries of successful attack resistance the required number of MSMs are also decreased. For example, in Fig \ref{prop1}, the required number of MSMs where total number of smart meters are $2000$ and malicious smart meters are $75\%$, is $16$ (green line). Whereas, the required number of MSMs is $11$ for the same scenario in Fig. \ref{prop5} and $9$ in Fig. \ref{prop10}. Each smart meter communicate with MSMs to send its noise data so in order to decrease the communication overhead appropriate number of MSMs can be chosen using this analysis.   \par

To access the performance of E-DPNCT for the collusion attack resistance, the results of a collusion attack on the E-DPNCT model are compared against the encryption based EPIC model. Less than $1\%$ data leak is considered as successful resistance against collusion attack. As shown in Fig. \ref{DPNCvsEPIC}, the number of required MSMs, on y-axis, are compared for different $\%$ percentage of malicious smart meters in an area is on x-axis. Results show that our E-DPNCT model required less number MSMs as compared to the EPIC privacy model \cite{epic}. For example, with $40\%$ malicious smart meters the required number of MSMs in EPIC is $11$ whereas, the required number of MSMs in E-DPNCT for the same is $6$.  

E-DPNCT has lower computational complexity and communication overhead as it does not require the sharing/communication of secret keys for data encryption and decryption. The EPIC model involves generating and sharing multiple secret keys for data encryption and decryption using homomorphic encryption which has high computation and communication cost as compared to E-DPNCT method. DP used in E-DPNCT utilises the random noise generation which has the minimal computation complexity. {Each smart meter generates a random number as part of differential privacy model and cost of generating a random number is $O(1)$. Whereas, aggregator aggregates $N$ masked readings and subtract aggregated noise from it which has computational complexity of $O(N)$ for load monitoring. For billing function the computational complexity is $O(T)$ where $T$ is billing period.}
E-DPNCT is also more fault tolerant in cases where smart meters fail to report their noise data. This is further discussed in section \ref{UitlityA} where the utility of E-DPNCT protected data is further evaluated. \par

\subsection{Privacy Analysis}\label{PrivacyA}

In order to verify the privacy and utility, we use the energy consumption data provided by Muratori \emph{et al.} \cite{data} to perform experiments to evaluate the accuracy and privacy of the E-DPNCT model. This simulated data provides the energy consumption data of $200$ households in watts with a granularity of $10$ minutes this gives us $6$ readings in an hour. The total number of readings received by a smart meter in a one month ($30 days$) billing period $T$, can be calculated as $T = 6 * 24 * 30 = 4,320$. We used the Numpy library of Python $3.0$ \cite{numpy} to implement E-DPNCT. To maintain simplicity while generating Laplacian noise, we fixed the $\epsilon = 1$. The impact of setting different values of privacy parameters $\epsilon$ have been explored previously by \cite{DPWGAN, PrivVsUtility} and are hence not elaborated in this paper. The point-wise sensitivity can be selected by each smart meter. As an example we experimented with $\Delta f $ = $max_{i,t} |x_{i,t}|$, $\Delta f $ = $avg_{i,t} |x_{i,t}|$ and half of both values in case of any outliers. We took the $mean = 0$ to measure the scale parameter $\lambda$. Generating a random number has the complexity cost of $ O(1)$ and our algorithm operates such that it adds a random number $n_{i,t}$ per reading $x_{i,t}$. \par
Previously, selecting a sensitivity parameter is not widely explored in differential private models for smart grids. Most common method of choosing sensitivity in literature is maximum value in the whole dataset \cite{Eibl2017,Hassan2020,dream}. Whereas, others choose the sum of electricity consumed by electric appliances in households \cite{Barbosa2016,PREEN}. In E-DPNCT, the impact of different sensitivity values(x-axis) on privacy (correlation coefficient) can be seen in Fig. \ref{fig:privacyvsSensitivity}. It can be seen in the figure that sensitivity has a direct impact on the privacy of data in DP as noise is calibrated according to the sensitivity of the query. The higher the sensitivity, the higher privacy is achieved. {We used correlation coefficient as privacy metric which tests the relationship between masked time series profile with original profile. } The correlation coefficient of original data with itself is $1$ showing no privacy which means that both data sets are the same whereas $\Delta f = max $ has the lowest correlation with original data showing highest privacy level.

\subsection{Utility Analysis} \label{UitlityA}
Mean Absolute Error (MAE) in total energy consumption for billing and load monitoring is calculated as follows~\cite{dream}:
\begin{equation}\label{Eq:5}
MAE = \sum \frac {|x_i - X_i|}{x_i}
\end{equation}
Where $x_i$ is the original energy consumption of the household and $X_i$ is the masked energy consumption of the household. \par


\subsubsection{Billing}
Calculation of bills is the first utility goal of our proposed model. The error in the billing period $T$ occurs due to noise added in the last $\Delta t$ as all the previous noise is cancelled out in subsequent $\Delta t$ where $\Delta t$ can be an hour, a day or a week. The error in the bill is reported by each house hold and it is cancelled in the next billing period as depicted by the  E-DPNCT Algorithm \ref{alg:DPNC1}. {The aggregator calculates bill using Block meter rate tariff \cite{tariff} where a consumer is charged with base unit price for first set of max allowed units and after that the excess units are charged with surcharge unit price as shown in Algorithm \ref{alg:DPNC3}. For experiments we set the $max allowed units$ =$2000kw$ for the billing period of one month. }\par
In Fig. \ref{fig:UtilityvsSensitivity} we compared the impact of sensitivity value over the accuracy in billing of a household using MAE. The E-DPNCT protected data using maximum load at an instant $t$ as sensitivity$(\Delta f)$ value exhibits largest error whereas selecting half of average load at an instant $t$ as sensitivity$(\Delta f)$ yields least MAE in billing results. Hence, robust selection of sensitivity value can be choice of energy consumers so that they can decide the level of privacy at the cost of accuracy.

{In fig. \ref{fig:maeDPNCT}, MAE in billing comparison between E-DPNCT and DRDP \cite{hassan2022} is depicted. It can be seen from the figure that E-DPNCT performs significantly better than DRDP.}

\subsubsection{Load Monitoring}
Calculation of total load in an area at an instant $(t)$ is the second utility goal in smart grids.  
Due to the use of infinite divisibility of Laplace noise, at each instant $t$, the aggregated masked load sent by all the smart meters in an area has privacy of $\epsilon_t $ as referred in \ref{PrelimSection}. In the ideal situation each MSM sends the aggregated noise to the aggregator to calculate accurate aggregated load. However, in situations where the aggregator does not receive any aggregated noise from MSMs the error would be $Lap(\lambda)$ (Theorem $24$). We evaluated mean absolute error in cases where smart meters could not report the added noise to MSMs. In fig. \ref{fig:lm}, the green line shows $10\%$ ($20$ out of $200$) smart meters fail to report the added noise back to the aggregator and the black line shows the original load in real time. It shows that even with $10\%$ smart meters not reporting the added noise information to their MSMs the load monitoring will be close to the original load monitoring curve. Further, Fig. \ref{fig:maelm} illustrates comparison of MAE in load monitoring in an area (y-axis) for different levels of faulty smart meters (x-axis), which fails to report their noise to the MSMs. From these results it can be deduced that the total mean absolute error (MAE) in load monitoring is only $0.1$ $kWh$ if $10\%$ smart meters do not send their noise information to the MSMs.   

\section{Conclusion and Future work}\label{Con}

In this paper, an E-DPNCT model with split noise distribution to multiple MSMs for a better resistance against privacy attacks is presented. We compared attack resistance of our E-DPNCT model with the state of the art privacy model EPIC. {In addition, the impact of sensitivity parameter on privacy and accuracy in billing and load monitoring is analysed. In conclusion, using multiple MSMs reduces the probability of a successful collusion attack in E-DPNCT and further preserve privacy and accuracy in billing and load monitoring. We also deduced that selecting sensitivity parameter for Laplace mechanism in differential privacy based privacy model plays crucial part in privacy vs. accuracy trade off. As part of future work, We plan on adding more utility functions on top of billing and load monitoring for example, time of use (ToU), value added services, etc. and access the impact of noise on them. }We also plan to work on detecting and mitigating data integrity attacks where adversary tries to inject false data for financial gains. The data perturbation privacy models for smart grids are easy target of such data integrity attacks. Hence, there is a need of a DP based privacy model which is resistant to data integrity attacks. 
\section{Data Availability}
The energy consumption data used in this paper is provided by Muratori \emph{et al.} \cite{data}. It is synthetically generated residential power consumption data of $200$ households.  





 





\bibliography{EDPNCT}



\section{Acknowledgements}
\vspace{-0.2cm}This publication has emanated from research conducted with the financial support of Science Foundation Ireland (SFI) and is funded under the Grant Number 18/CRT/6222. This work additionally received support from the Higher Education Authority (HEA) under the Human Capital Initiative-Pillar 3 project, Cyberskills.\vspace{-0.2cm}

\section{Author contributions statement}
K.H. presented the idea which is enhanced and refined by D.O and M.H.R. K.H. planned and conducted the experiments. K.H, D.O, and M.H.R performed the data analysis. K.H wrote the initial draft of the manuscript which is reviewed and edited by D.O. and M.H.R. D.O. also participated in the write-up of the manuscript.  The manuscript is reviewed by T.N in the end. All authors reviewed the data and the manuscript and agreed to the published version of the manuscript.

\section{Additional Information}
\textbf{Competing interests:} The authors declare no competing interests.
\textbf{Corresponding author:} Correspondence and requests for materials should be addressed to K.H.

\end{document}